# Uncertainty Evaluation of the Caesium Fountain Primary Frequency Standard NIM6

**Fasong Zheng[1,2], Weiliang Chen[1,2], Kun Liu[1,2], Shaoyang Dai[1,2], Nianfeng Liu[1,2], Yuzhuo Wang[1,2] and Fang Fang[1,2]**

[1] National Institute of Metrology (NIM), Beijing 100029, China
[2] Key Laboratory of Time and Frequency and Gravity Acceleration of State Administration for Market Regulation, Beijing 100029, China

E-mail: fangf@nim.ac.cn

## Abstract

A new caesium (Cs) fountain clock NIM6 has been developed at the National Institute of Metrology (NIM) in China, for which a comprehensive uncertainty evaluation is presented. A three-dimensional magneto-optical trap (3D MOT) loading optical molasses is employed to obtain more cold atoms rapidly and efficiently with a tunable, uniform density distribution. A heat pipe surrounding the flight tube maintains a consistent and stable temperature within the interrogation region. Additionally, a Ramsey cavity with four azimuthally distribution feeds is utilized to mitigate distributed cavity phase shifts. The Cs fountain clock NIM6 achieves a short-term stability of $1.0 \times 10^{-13} \tau^{-1/2}$ at high atomic density, and a typical overall fractional type-B uncertainty is estimated to be $2.3 \times 10^{-16}$. Comparisons of frequency between the Cs fountain NIM6 and other Cs fountain Primary Frequency Standards (PFSs) through Coordinated Universal Time (UTC) have demonstrated an agreement within the stated uncertainties.

## 1. Introduction

Caesium fountain Primary Frequency Standards (PFSs) have demonstrated the best realization of the definition of the second in the International System of Units (SI) for nearly three decades. Since the invention of the first laser-cooled Cs fountain clock [1], Cs fountain PFSs have been developed worldwide [2-18]. To date, more than ten Cs fountain PFSs have reported frequency uncertainties ranging from a few parts in $10^{15}$ to a few parts in $10^{16}$ to the International Bureau of Weights and Measures (BIPM), contributing to the calibration of International Atomic Time (TAI) [19]. Among these, the Cs fountain clock NIM5, developed at the NIM in China, has been reporting monthly data to the BIPM since 2014 [12]. It operates with a typical fractional frequency instability of $1.5 \times 10^{-13} (\tau/s)^{-1/2}$ and a type-B uncertainty of

$6.8 \times 10^{-16}$ now, with microwave-related frequency shifts being the predominant source of uncertainty.

A new Cs fountain PFS, the Cs fountain NIM6 has been constructed and operated now [20-26]. The goal was to achieve lower uncertainty by implementing several improvements over the Cs fountain NIM5. The Cs fountain NIM6 features a three-dimensional magneto-optical trap (3D-MOT) loading optical molasses is used to obtain more cold atoms rapidly and efficiently with a tunable, uniform density distribution. Additionally, a heat pipe surrounds the flight tube to maintain a more uniform and stable temperature within the interrogation region. Moreover, a Ramsey cavity with four azimuthally distribution feeds has been employed to mitigate the effects of distributed cavity phase shifts.

In this paper, we report the first comprehensive uncertainty evaluation of the Cs fountain NIM6. Section 2

describes the physics package, the optical system, and the microwave synthesis. Section 3 gives the details of the operation. Section 4 evaluates the frequency biases and associated uncertainties for each physical effect. Section 5 discusses the results from the frequency comparisons between the Cs fountain NIM6 and other Cs fountain PFSs through UTC. Finally, section 6 gives the conclusions.

## 2. Description of the apparatus design

The designs, operating conditions and preliminary results of the Cs fountain NIM6 have been presented and discussed in References [20-24]. In this section, we will focus exclusively on the major features.

### 2.1 Physical package

A significant difference of the NIM6 over the NIM5 is the method of cold atom collection. In the NIM6, atoms are initially collected in a lower 3D-MOT chamber and subsequently launched to an upper optical molasses (OM) chamber at a 10° angle from vertical, as illustrated in figure 1. Both the MOT and OM chambers employ the (1, 1, 1) cooling beam geometry, with laser lights delivered through optical fibers. The atomic density at the OM stage is fine-tuned by adjusting the laser detuning during the post-cooling phase following the MOT stage. This approach not only enables quicker collection of cold atoms compared to the direct molasses method used by the NIM5 but also achieves a more uniform density distribution compared to loading molasses with a 2D-MOT. The 10° launch angle serves to prevent background Cs atoms from entering the detection chamber, thereby reducing detection noise. Additionally, a $TE_{011}$-mode state-selection microwave cavity, measuring 21.50 mm in inner diameter and 42.96 mm in length, is installed at the upper portion of the OM chamber.

The detection chamber is positioned above the OM chamber, adhering to a design similar to that of the NIM5 [12]. In this configuration, two detection beams are spaced 30 mm apart vertically, facilitating the spatially distinct detection of atoms in the $^6S_{1/2}$ $F$ =3 (lower zone) and $^6S_{1/2}$ $F$ =4 (upper zone) ground states. Three pairs of large square coils are strategically placed at the lower part of the physical package, arranged orthogonally to each other. Additionally, an extra coil is installed at the top of the detection chamber (near the bottom endcap of the magnetic shields). This setup is designed to compensate for environmental magnetic fields and to provide a relatively uniform B field, serving as the quantization axis following OM cooling.

The Ramsey interrogation region is comprised primarily of a flight tube with a Ramsey cavity inside, and an ion-getter pump on the top. The pump was actived during the vacuum baking and disconnected during the fountain operation to ensure that the physical package ground (GND) connection is made solely through the microwave feedthroughs. To maintain a uniform and stable temperature along the atomic trajectory, the flight tube is encased by an isothermal liner heat pipe. PT100 temperature sensors are affixed to the exterior wall of the flight tube, with further details available in Reference [20]. Constructed from oxygen-free copper, the $TE_{011}$-mode Ramsey cavity measures 24.20 mm in diameter and 28.62 mm in height. It features four equatorially placed feeds, with opposing feeds designated as the $X/Y$ axis [24]. All connections to the cavity are sealed with indiums to prevent microwaves leakage into the atomic trajectory. Above the cavity, a copper tube with an 18 mm inner diameter is installed to further block the microwaves leaking into the interrogation region. This cavity's loaded quality factor is approximately 12000 and its resonant frequency matches the Cs clock frequency within 10 kHz at room temperature. The interrogation region is then enclosed by three cylindrical magnetic shields made of μ-metal. A double-layer solenoid coil is cross-wrapped with two additional coils at the two ends to provide a uniformly distributed magnetic field (C-field).

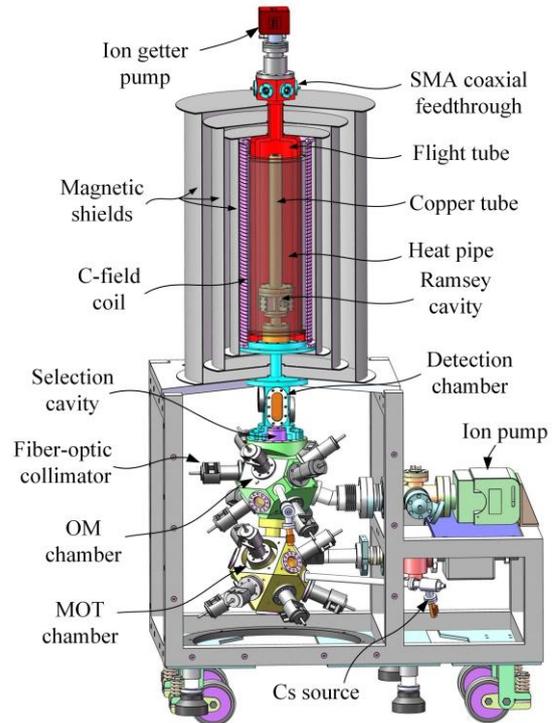

**Figure 1.** Diagram of the physics package of the Cs fountain NIM6. MOT, magneto-optical trap; OM, optical molasses.

### 2.2. Laser set-up

Cs atoms are initially collected in the lower 3D-MOT chamber and subsequently launched toward the center of the OM chamber. The separation between these two centres is 280 mm, and the flying time of the atomic cloud is about 50 ms with a launching velocity of 5.5 m/s. To minimize the collisional shift, the atomic temperature after MOT is adjusted by tuning the intensities and frequency detunings of the post-cooling beams to make sure the cloud expanded enough when it reaches the OM chamber's center. Within the OM chamber, the velocity and temperature of the atomic cloud are readjusted, after which the cloud is vertically launched into the flight tube. This design offers several advantages: it captures cold atoms more rapidly compared to the direct optical molasses approach used in the Cs fountain clock NIM5, it reduces the presence of background Cs gas in the detection chamber through a differential pumping, and it achieves a more uniform atom density distribution than that obtained from loading OM from a 2D-MOT. A notable feature of this setup is that the cooling beams for MOT and OM are applied sequentially rather than simultaneously, allowing a single laser to serve both functions.

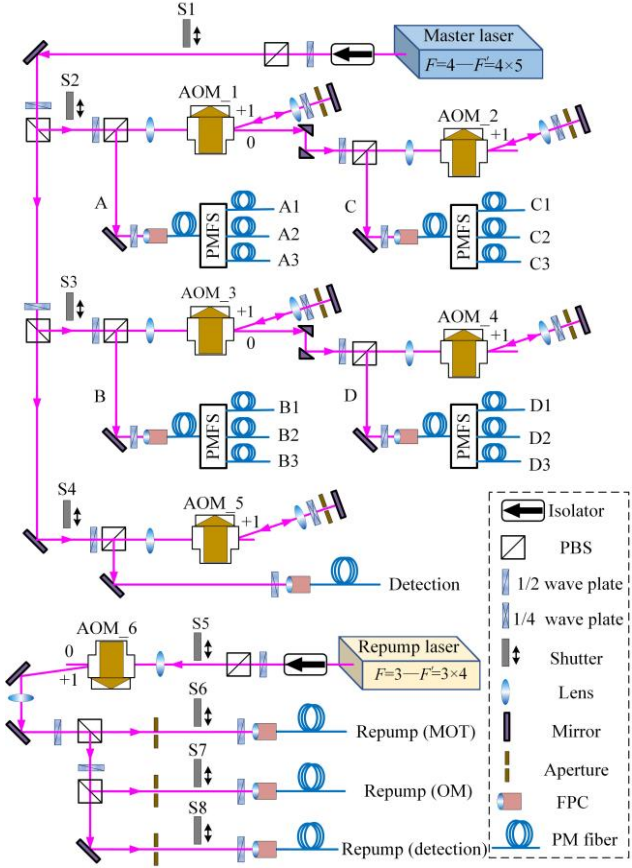

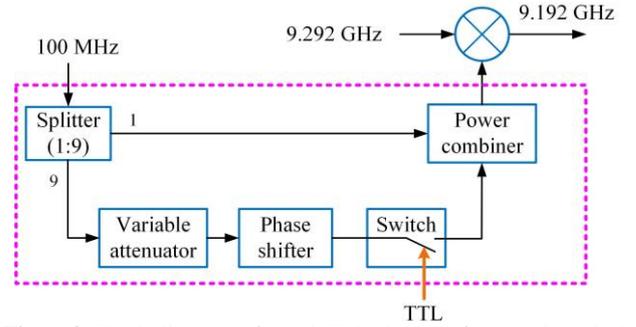

**Figure 2.** Diagram of the optical setup. PBS, polarizing beam splitter; PM fiber, polarization-maintaining fiber; FPC, fiber port coupler; PMFS, polarization-maintaining fiber splitter; AOM, acousto-optic modulator.

The optical system layout is depicted in figure 2. A single 852 nm tapered amplifier (TA) with an output power of approximately 800 mW provides the laser lights for both optical traps (MOT and OM) and the detection system. The laser frequency is stabilized at the crossover peak between $F=4$ to $F'=4$ and $F'=5$ of the saturated absorption spectrum. This frequency-locked beam is divided into three paths. The first two paths are each frequency-shifted by making a double-pass through an acousto-optic modulator (AOM), which generates the upward (A1-A3) and downward (B1-B3) cooling beams for the MOT. The zero-order beams from these two AOMs are frequency-shifted by making a further double-pass through another two AOMs respectively, producing the cooling beams (C1-C3 and D1-D3) for the OM. The third beam is similarly frequency-shifted by a double-pass through an AOM for the detection. The detection beam is then split into two branches to separately detect atoms in the $F=4$ and $F=3$ states. To enhance the signal-to-noise ratio, these beams are retroreflected to create standing waves. Additionally, the reflection from the lower part of the upper beam is obstructed to push away any residual atoms in the $F=4$ states during the atoms ascending and descending.

An additional 852 nm laser with an output power of approximately 150 mW is frequency-locked at the crossover peak between $F=3$ to $F'=3$ and $F'=4$ of the Cs saturated absorption spectrum for repumping. The beam frequency is shifted by making a single-pass through an AOM. Subsequently, the beam is divided into three separate beams. These beams are used as the re-pumping beams for both the MOT and OM, as well as for the re-pumping in the lower detection zone to specifically detect atoms in the $F=3$ state.

### 2.3. Microwave synthesis

The fountain clock NIM6 employs microwave synthesizers similar to those used in the fountain NIM5 [12], with a notable enhancement in its interrogation microwave synthesizer. This component is equipped with an improved 100 MHz Mach-Zehnder interferometric switch, the architecture of which is illustrated in Figure 3. The major modification in this version involves the removal of a power amplifier from one of the interferometric arms, significantly reducing the related phase noise. As detailed in section 4.2.3, this switch introduces a negligible transient phase in the transmitted field after it is switched on.

An active hydrogen maser (BIPM code 1404821) with an excellent frequency stability of about 4.5E-14 at 1 s is used as a local oscillator. Its 100 MHz output signal is used to phase-lock a dielectric resonator oscillator (DRO), which in turn generates a 9.3 GHz signal. This 9.3 GHz signal is mixed with a 7.368 MHz signal from a computer-controlled synthesizer first and then mixed with a 100 MHz signal from the Mach-Zehnder interferometric switch, to generate a 9.192

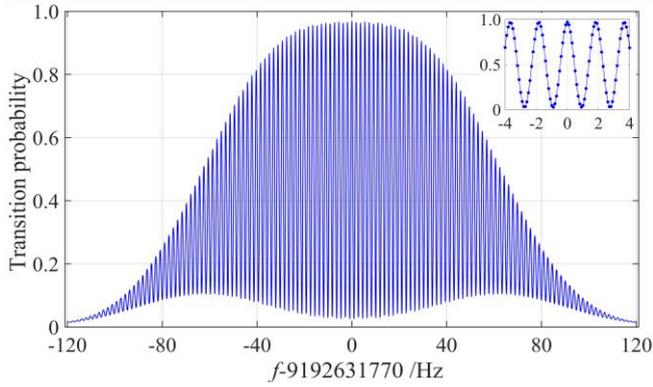

**Figure 4.** Ramsey fringe of the $|3, 0\rangle \leftrightarrow |4, 0\rangle$ clock transition in NIM6. The central fringe exhibits a FWHM width of 0.92 Hz and a contrast of 95%.

GHz signal that drives the clock transitions. The switch is regularly proven to attenuate the 9.192 GHz signal by more than 60 dB. Finally, the 9.192 GHz signal is distributed to 4 feeds of the Ramsey cavity via a 1-to-4 microwave splitter unit.

## 3. Operation of the primary frequency standard

The operation cycle of the NIM6 is similar to that of the NIM5, with the primary difference being the collection and cooling of atoms in the lower MOT chamber, followed by their launch to the OM chamber. In the OM stage, the atoms are re-adjusted to a vertical launching velocity of 4.1 m/s and further cooled to approximately 2 μK. After the OM launch, a microwave π-pulse is applied for state selection, transferring atoms from $|F=4, m_F=0\rangle$ state to $|F=3, m_F=0\rangle$ state. Atoms remaining in the $|F=4\rangle$ manifold are then pushed away by a resonant light pulse. The atoms in the clock state $|F=3, m_F=0\rangle$ pass through the Ramsey cavity twice during their ballistic flight before entering the detection chamber, where the transition probability is measured to lock the microwave frequency to the Cs clock transition [12, 16]. This process yields a typical Ramsey fringe pattern, as shown in figure 4. The full width at half maximum (FWHM) width of the central Ramsey fringe measures 0.92 Hz, with a contrast of approximately 95%.

## 4. Systematic biases and uncertainty evaluation

### 4.1 Type-A uncertainty evaluations

The Allan deviation of the relative frequency (measured by the NIM6 against the H-maser), extrapolated to zero density, serves as the type-A uncertainty.

The measured relative frequency is shifted by the collisions between cold atoms, which is proportional to atomic density (or atom number). In a routine operation, the NIM6 alternates between high and low atomic density conditions, as detailed in section 4.2.2. This approach allows for the extrapolation of relative frequency to zero density.

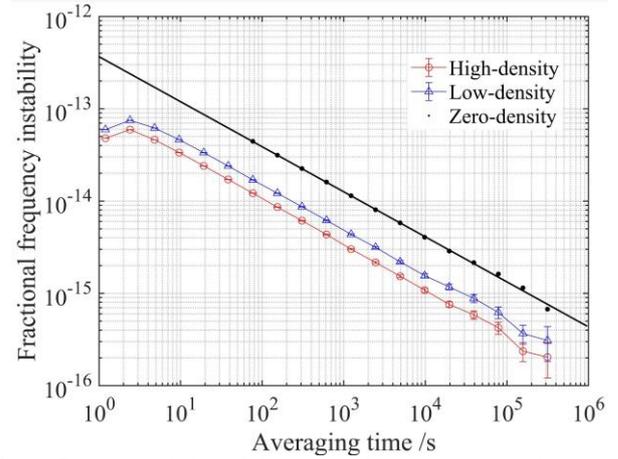

**Figure 5.** Allan deviation of the relative frequency measured by NIM6 against hydrogen maser H-21 over 33 days, from MJD 60463 to MJD 60495. MJD: Modified Julian Day. Open circles, open triangles, and solid circles represent the instabilities of relative frequency at high, low and zero atomic densities respectively. Error bars indicate one-standard-deviation confidence levels. The solid black line represents the fit for the extrapolated relative frequency at zero atomic density.

During a measurement campaign, the relative frequencies $f_H$ and $f_L$, along with the detected atomic numbers $N_H$ and $N_L$ at high and low atomic densities respectively, are recorded and analyzed. The relative frequency extrapolated to zero density is given by

$$f_{Zero} = \frac{k f_L - f_H}{k-1} \tag{1}$$

where $k=N_H/N_L$. Frequent alternations between two densities allow for the neglection of reference maser instabilities as common mode fluctuations. Utilizing equation (1), the uncertainty of the extrapolated frequency $f_{Zero}$ is [27]

$$\sigma_Z(\tau_L + \tau_H) = \frac{1}{k-1}\sqrt{k^2 \sigma_L^2(\tau_L) + \sigma_H^2(\tau_H)} \tag{2}$$

where $\sigma_H(\tau_H)$ ($\sigma_L(\tau_L)$) represent the Allan deviation of the relative frequency instability at the high (low) density, and $\tau_H$ ($\tau_L$) is the averaging time. The atomic density is adjusted without altering the duration of the fountain cycle. Hence, $\tau_H=\tau_L=\tau$.

The fountain clock NIM6 is alternately operated at two different atomic densities with $k \approx 2.3$. Figure 5 displays the typical relative frequency instabilities against H-21 master over an averaging period of 33 days. The Allan deviations for measured relative frequencies at high, low and zero atomic densities are indicated by open circles, open triangles and solid circles, respectively. The short-term stability values are $\sigma_H=1.0\times10^{-13}(\tau/s)^{-1/2}$, $\sigma_L=1.4\times10^{-13}(\tau/s)^{-1/2}$, and $\sigma_Z=3.7\times10^{-13}(\tau/s)^{-1/2}$. The NIM6 uses $\sigma_Z$ as the type-A uncertainty during frequency measurement. It decreases with measurement time, 25-day and 30-day measurements yield uncertainties of $2.5\times10^{-16}$ and $2.3\times10^{-16}$, respectively.

### 4.2 Type-B uncertainty evaluations

The output frequency of a primary frequency standard is subjected to systematic biases arising from various physical effects. The subsequent sections detail the measurement of these frequency biases and the evaluations of the associated uncertainties.

### 4.2.1 Second-order Zeeman shift

The C-field provides the quantization axis and meanwhile induces a second-order Zeeman shift in the clock transition frequency between $|F=3, m_F=0\rangle$ and $|F=4, m_F=0\rangle$. This shift is the largest systematic frequency shift of the NIM6, as in many other atomic fountains. A measurement approach similar to that used in the fountain NIM5 is employed to quantify this shift [12]. Atoms are launched to various heights with an increment of 10 mm in each step. The central Ramsey fringes of the magnetic field-sensitive transition $|F=3, m_F=1\rangle \leftrightarrow |F=4, m_F=1\rangle$ are tracked and recorded unambiguously. Figure 6 displays the Ramsey fringes for this sensitive transition, observed at a routine launching height of 850 mm above the OM chamber center. From these measurements, a time-averaged $B$ fields $\langle B_C(h)\rangle$ at each height $h$ can be obtained from the following formula with an unit in nT

$$f_{4,1-3,1}(h) = f_0 + 7.0083 \cdot \langle B_C(h)\rangle \quad (3)$$

where $f_0$= 9 192 631 770 Hz represents the clock transition frequency, $f_{4,1-3,1}(h)$ denotes the central Ramsey fringe frequency of the $|F=3, m_F=1\rangle \leftrightarrow |F=4, m_F=1\rangle$ transition at each height $h$. By launching atoms to various heights, a magnetic field map is constructed through deconvolution of the time-averaged magnetic fields and the ballistic times of flight at different heights, as shown in figure 7. It provides a detailed spatial profile of the magnetic field strength throughout the interrogation path of the atoms.

The fractional frequency shift due to second-order Zeeman effect can be obtained from the Breit–Rabi formula as [28]

$$\left(\frac{\Delta f}{f_0}\right)_{\text{Zeeman}} = \frac{(g_J - g_I)^2 \mu_B^2}{2h^2 f_0^2}\langle B_C^2\rangle \quad (4)$$

where $g_J$, $g_I$ are Lande $g$-factors with and without the nuclear spin respectively, and $\mu_B$ is the Bohr magnetron. Usually, the time averaged field $\langle B_C\rangle^2$ is measured and used to calculate the second-order Zeeman shift. When the C-field is inhomogeneous, $\langle B_C^2\rangle$ is given by $\langle B_C^2\rangle = \langle B_C\rangle^2 + \sigma^2$ where $\sigma^2$ represents the variance of $B$ along the atomic trajectory. The term $\sigma$ indicates the C-field's uniformity, leading to an uncertainty in the evaluation of the frequency bias. The total uncertainty of the second-order Zeeman frequency shift is

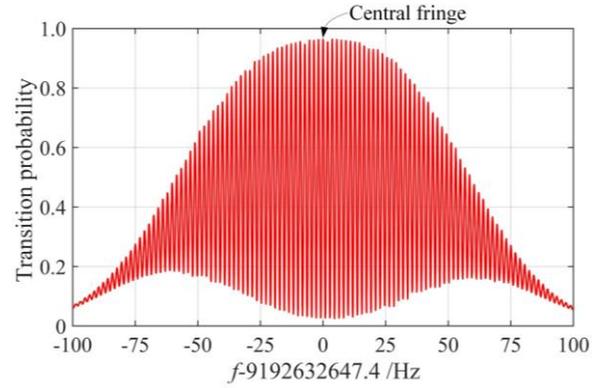

**Figure 6.** Ramsey fringes for the $|3, 1\rangle$ to $|4, 1\rangle$ transition observed during normal operation at a launching height of 850 mm, with a frequency scanning step of 0.1 Hz.

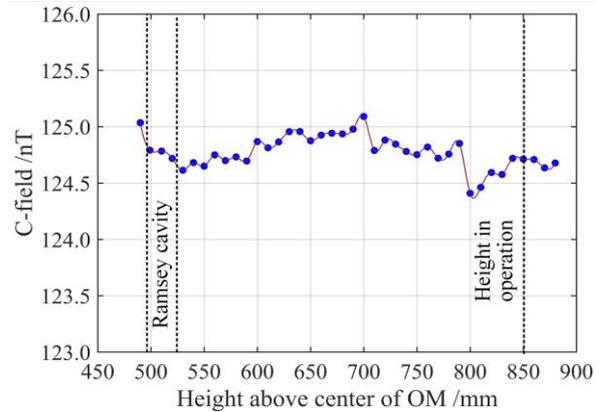

**Figure 7.** C-field map above the center of the OM chamber.

related to this term and the C-field temporal fluctuation. It can be expressed as

$$\delta\left(\frac{\Delta f}{f_0}\right)_{\text{Zeeman}} = \frac{(g_J - g_I)^2 \mu_B^2}{2h^2 f_0^2}\sqrt{\left(2\langle B_C\rangle\frac{d\langle B_C\rangle}{dt}\right)^2 + (\sigma^2)^2} \quad (5)$$

where $d\langle B_C\rangle/dt$ denotes the temporal fluctuation of the mean C-field strength. This factor is critical as it contributes to the overall uncertainty in the evaluation of the second-order Zeeman frequency shift.

In routine operations of the NIM6, a time-averaged magnetic field is measured to be 125.2 nT, corresponding to a calculated relative second-order Zeeman frequency shift of $72.88 \times 10^{-15}$ from equation (4). Figure 7 shows that the inhomogeneity of the C-field within the interrogation region is less than 0.7 nT, leading to a calculated fractional frequency uncertainty of $2.3 \times 10^{-18}$ from equation (5). The temporal fluctuation of the C-field introduces a larger uncertainty to the second-order Zeeman frequency shift. This fluctuation is monitored by tracking the variation in the central Ramsey fringe position of the $|F=3, m_F=1\rangle \leftrightarrow |F=4, m_F=1\rangle$ transition. As shown in figure 8, a monthly variation of less than 0.4 Hz was observed, corresponding to an uncertainty in the second-order Zeeman shift correction of approximately $6.6\times10^{-17}$, as calculated from equation (5). Consequently, the relative frequency shift due to the second-order Zeeman effect is evaluated to be $72.88\times10^{-15}$ with an associated uncertainty of $0.7\times10^{-16}$.

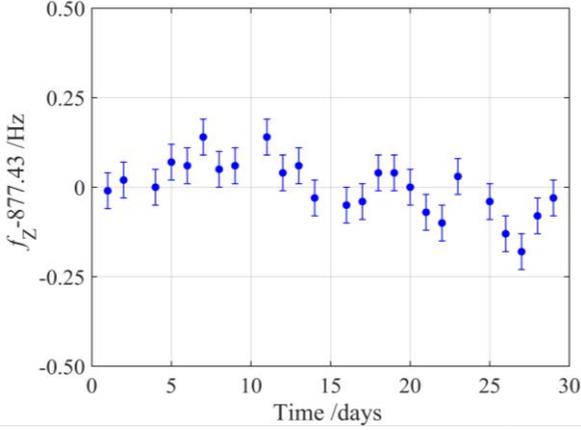

**Figure 8.** Tracking of the central frequency of the $|3, 1\rangle$ to $|4, 1\rangle$ transition. $f_Z$ represents the Zeeman frequency.

### 4.2.2 Collisional shift

Collisions among cold atoms during their ballistic flight above the Ramsey cavity result in a systematic shift in the clock frequency. This shift is proportional to the atomic density [29].

To measure the magnitude of the cold collisional shift, the fountain NIM6 operates alternatively between high and low atomic densities (or atom number) by adjusting the power of the state selection microwave pulse. As described in section 4.1, the relative frequencies $f_H$ and $f_L$, along with the detected atomic numbers $N_H$ and $N_L$ at the high and low densities respectively, are systematically recorded. Using equation (1), the frequency at zero density is extrapolated.

At low atomic density, the collisional shift can be expressed as

$$f_{C-L} = f_H - f_{Zero} = \frac{f_H - f_L}{k - 1} \qquad (6)$$

As explained in Reference [12], the type-B uncertainty induced by cold collisions can be expressed as

$$\begin{cases} \delta\left(\dfrac{f_{C-L}}{f_0}\right)_{\text{collision}} = \dfrac{1}{f_0}\left(\dfrac{f_H - f_L}{(k-1)^2}\right)\sigma_k \\ \sigma_k^2 = \sigma_{\text{nonlinear}}^2 + \left[\left(\dfrac{\delta N_L}{N_L}\right)^2 + \left(\dfrac{\delta N_H}{N_H}\right)^2\right]k^2 \end{cases} \qquad (7)$$

where $\sigma_k$ is the uncertainty of the ratio $k$, $\sigma_{\text{nonlinear}}$ is the nonlinearity between the measured atom numbers and the average densities, as stated in [30, 31]. $\delta N_L$, $\delta N_H$ are the atom number uncertainties. In practice, the measured fluorescence signals are proportional to atom numbers rather than atomic densities. This nonlinearity is checked by calculating the atomic spatial profile difference between high and low atom numbers to estimate the $\sigma_{\text{nonlinear}}$. To investigate this, the time-of-flight (TOF) profiles for atoms at low and high atomic densities during the atoms' descent are measured, as shown in figure 9. The profiles and the linewidths of TOF signals at the two atomic densities are almost identical, with a variance of less than 1%.

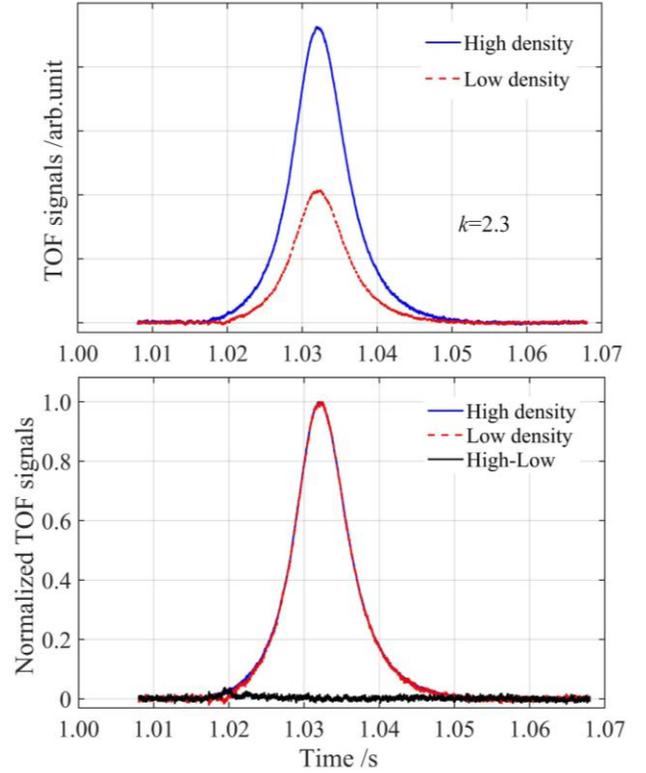

**Figure 9.** Time-of-flight (TOF) signals measured during the atoms' descent. The black line is the difference between the normalized signals for high and low atomic densities, with $k$=2.3. The line shapes of TOF signals at different densities are almost identical.

In the NIM6 routine operations, the ratio $k$ between high and low densities is approximately 2.3. Atom number fluctuations, both daily and from day to day, are less than 4%. $\sigma_{\text{nonlinear}}$ is estimated to be 0.01, and then the $\sigma_k$ is evaluated to be 0.13 from equation (7). Consequently, the fractional frequency shift due to cold collisions at low density is usually about $-2.2\times10^{-15}$ (25-day measurement) with an uncertainty of $1.7\times10^{-16}$.

### 4.2.3 Microwave-power-related frequency shifts

The frequency shifts induced by microwave mainly include the microwave leakage-induced frequency shift, the distributed cavity phase (DCP) frequency shift, and the spectral impurities in the Ramsey interrogation microwave signals-induced frequency shift.

*A. Microwave leakage.* Microwaves leaking from the cavity, frequency synthesizer, and microwave connections can interact with atoms, causing frequency shifts [32]. Atoms are susceptible to interactions with these leakage microwave fields at three distinct stages: between state-selection and the first Ramsey pulse, between the two Ramsey pulses, and between the second Ramsey pulse and the detection [33]. To minimize these effects in the Cs fountain NIM6, all microwave connections inside the vacuum are carefully sealed with indium, reducing potential leakage. Additionally, a 40 cm long copper tube with an inner diameter of 18 mm (shown in figure 1) is installed above the Ramsey cavity to further prevent the interrogation region from unwanted microwaves. This design theoretically attenuates external microwave fields by more than 30 dB at the apogee of the typical launching height of 850 mm.

To further mitigate the possibility of microwave leakages, a Mach-Zehnder interferometric switch (illustrated in figure 3) with a measured attenuation factor of about 60 dB at 9.192 GHz is employed. This switch can be turned off the 9.192 GHz signal when the atoms are outside the Ramsey cavity. The effects of microwave leakage were evaluated by measuring the relative frequency difference between the switch on and off during the Ramsey interrogation time. Over a measurement duration of 5 days, the relative frequency differences were found to be $(-3.5 \pm 5.0) \times 10^{-16}$ for a Ramsey power $\pi/2$ pulse and $(3.2 \pm 7.0) \times 10^{-16}$ for a $3\pi/2$ pulse, indicating that the frequency shift is less than $1.0 \times 10^{-17}$.

A Triggered-Phase Transient Analyzer (TPTA) was used to examine the transient phase induced by the interferometric switch, similar to the procedure followed for NIM5 [12, 21]. Figure 10 represents the measurement results, showing an averaged phase difference between the two Ramsey pulses of approximately 0.033 µrad over a measurement duration of 172700 s (about 2 days). This corresponds to a relative frequency shift of $0.6 \times 10^{-18}$. The measurement is constrained by the stability of the microwave signal, and the uncertainty in the frequency bias induced by the interferometric switch is estimated to be below $1.0 \times 10^{-16}$.

*B. DCP.* The spatial variation of the microwave phase across the apertures in the cylindrical Ramsey cavity can lead to a phase difference between the two microwave pulses encountered by atoms as they pass through the cavity at different transverse positions during their upward and downward trajectories. This effect results in a significant shift known as the DCP shift [34-38].

As described in References [35, 36], the transverse phase variations can be decomposed into a Fourier series $\cos(m\varphi)$ in cylindrical coordinates $(\rho, \varphi, z)$, where $m$ is an integer mode number and $\varphi$ is the azimuthal angle. To the lowest order, the transverse phase of the cavity field is proportional to $\rho^m \cos(m\varphi)$. Since the cavity apertures restrict atomic trajectories to small $\rho$, only terms with $m \leq 2$ contribute significantly to the frequency shift [36, 37]. The $m=0$ phase variations are mainly caused by power flow from the feeds at the cavity midplane to the endcaps. This power flow implies very small transverse phase variations and large longitudinal phase variations, primarily due to the endcap losses. The $m=1$ phase variations primarily arise from power flow across the cavity, potentially due to imbalances between opposing feeds or inhomogeneities in the resistance of the copper cavity walls. Near the cavity axis, this $m=1$ phase variation presents as linear gradient. The $m=2$ phase variations mainly result from transverse power flow from the feeds to the cavity walls. Non-uniform state detection, particularly along the feed axis of the Ramsey microwaves, as well as an offset of the atomic cloud from the centre of the Ramsey cavity can lead to a $m=2$ DCP shift [36, 38].

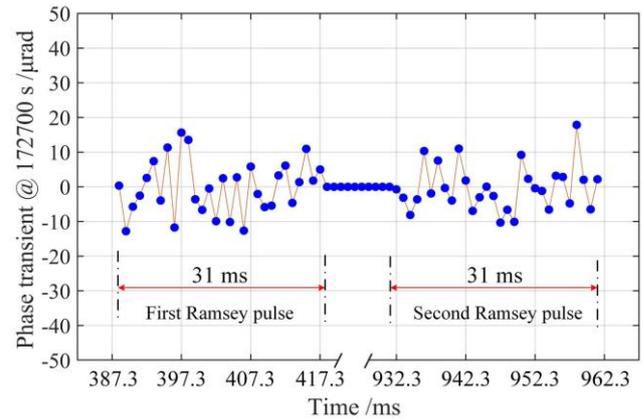

**Figure 10.** Averaged phase difference between two Ramsey pulses measured during NIM6 operation using the Mach-Zehnder interferometric switch. The averaged phase difference is about 0.033 µrad, recorded over a measurement duration of 172700 s (about 2 days).

Typically, the $m=0$ and 2 phase variations introduce sufficiently small DCP frequency shifts, which can be accurately calculated by considering the cavity's geometry, and the dimensions and positions of the atomic cloud in the cavity when they passing by. The NIM6 Ramsey cavity has similar dimensions to that of the PTB-CSF2 and made of same material [5]. They are both cylindrical TE$_{011}$ cavities constructed from OFHC copper, with an inner diameter of 48.4 mm, a height of 28.6 mm, and top and bottom endcaps with 10 mm diameter apertures. Moreover, the initial atomic cloud radius at launch and the atomic temperature are also very close. Consequently, according to the DCP evaluation in the PTB-CSF2, as detailed in [37], the $m=0$ DCP frequency shift in the NIM6 is estimated to be less than $1.0 \times 10^{-17}$ at optimum microwave amplitude (corresponding to $\pi/2$ Ramsey pulse), with this value treated as its uncertainty estimate. In the NIM6, a Ramsey cavity with four azimuthally distributed feeds, as well as a 45° angle between the probe light propagation axis and the microwave feed direction are utilized. These measures, as analysed in [34, 36, 38], effectively mitigates the $m=2$ DCP frequency shift to a negligible level.

However, the $m=1$ transverse phase variations can produce a significant DCP frequency shift if the mean transverse positions of the atomic cloud in the cavity differ between two passages (upward and downward). Given the linear gradient of the $m$=1 phase variation, these related DCP frequency shifts can be experimentally evaluated by tilting the entire physical package to intentionally increase the mean transverse displacement of the atomic cloud during the two cavity passages. When the fountain clock is tilted by a small angle $\theta$, the relative frequency shift is proportional to $\theta$ and can be described as

$$\left(\frac{\Delta f}{f_0}\right)_{DCP} = \gamma\theta \qquad (8)$$

where $\gamma$ represents the sensitivity coefficient of the frequency shift to the tilt angle $\theta$. The uncertainty in the $m$=1 DCP shift is mainly determined by two main factors: the accuracy with which $\theta$ can be adjusted to achieve the optimal tilt, and the uncertainty of the $\gamma$ value.

The Ramsey cavity of the Cs fountain NIM6 has four azimuthally distributed feeds located on the equatorial plane. The axis formed by two face-to-face feeds is designated as the $X$-direction of the fountain, while the axis formed by the other pair of feeds is labeled as the $Y$-direction. To minimize the DCP frequency shift, the phase and amplitude of the microwaves fed into the cavity's four feeds can be carefully balanced. This is achieved by individually adjusting external attenuators and phase.

To evaluate the $m$=1 DCP frequency shift along the $X$ direction, the Cs fountain NIM6 was tilted along the $X$ axis while keeping the $Y$ direction fixed. The relative frequency differences between two opposite feeding directions ($\varphi$=0 and $\varphi$=$\pi$) were measured as a function of the tilt angles under two different Ramsey pulses ($\pi/2$ and $3\pi/2$). The results are shown in figure 11(a). The optimal tilt in the $X$ direction where the frequency difference is minimized, was determined for each Ramsey pulse through linear fitting of the measured data using the least-squares method. Following this, the fountain NIM6 was then tilted along the $Y$ direction while the $X$ direction was set to the optimal tilt point determined in figure 11(a), and the results are shown in figure 11(b).

Form the measured data presented in figure 11, in the $X$ direction, an optimal tilt angle of -0.03 mrad was obtained with the $\pi/2$ Ramsey pulse. Considering the angle adjustment error of 0.05 mrad, the optimal angle has an uncertainty of 0.11 mrad, which was estimated employing the Monte Carlo method. To validate this, the optimal tilt angle obtained from the $3\pi/2$ Ramsey pulses was also considered, revealing a difference of 0.09 mrad between the two estimates. Analogously, in the $Y$ direction, an optimal tilt angle of -0.11 mrad was obtained for the $\pi/2$ Ramsey pulse, accompanied by a fitting uncertainty of 0.44 mrad. The comparison between the optimal angles derived from different Ramsey pulses ($\pi/2$ and $3\pi/2$) in the $Y$ direction showed a discrepancy of 0.38 mrad. Consequently, the uncertainties associated with the optimal tilt angles are quantified as 0.11 mrad in the $X$ direction and 0.44 mrad in the $Y$ direction.

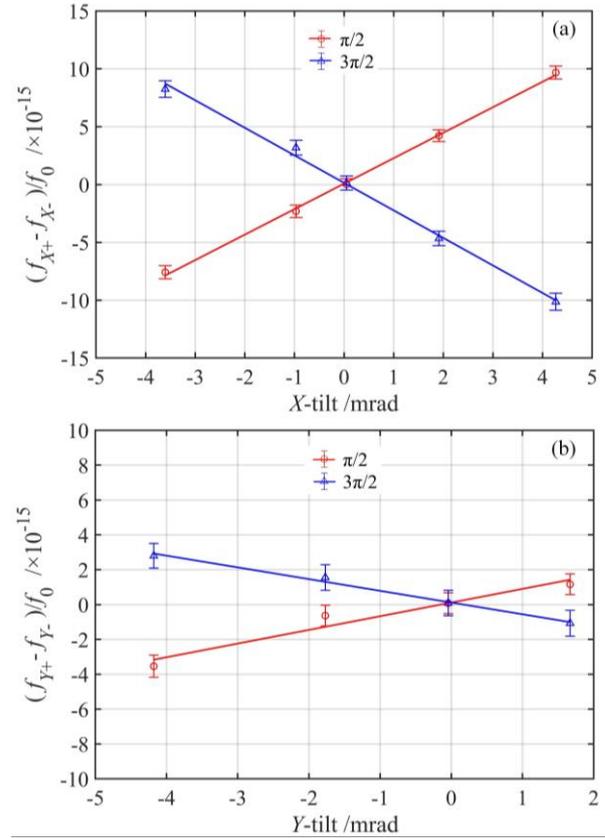

**Figure 11**. Measured fractional frequency differences between feeding at $\varphi$=0 and $\pi$ as a function of tilt angles along the cavity feed directions $X$ and $Y$. (a) $X$ direction; (b) $Y$ direction. At the optimal tilt angle, the fractional frequency difference between the two opposite feeds is minimized. The red circles represent measurements with $\pi/2$ Ramsey pulse, while the blue triangles correspond to $3\pi/2$ Ramsey pulses. The lines show a linear fit of the data, yielding slopes of $2.20\times10^{-15}$mrad$^{-1}$ and -2.38$\times10^{-15}$ mrad$^{-1}$ in the $X$ direction, and $0.78\times10^{-15}$mrad$^{-1}$ and -0.67$\times10^{-15}$ mrad$^{-1}$ in the $Y$ direction, for $\pi/2$ and $3\pi/2$ Ramsey pulses, respectively.

From the fitted curves in figure 11, the measured tilt sensitivity at the optimum power ($\pi/2$) in the $X$ direction is $2.20\times10^{-15}$ mrad$^{-1}$. Symmetric feeding of the Ramsey cavity significantly reduces the phase gradient, thereby lowering the sensitivity of the frequency shift. Assuming the microwave power from two feeds is balanced within 20%. For the optimal tilt angle with uncertainty of 0.11 mrad, the fractional frequency uncertainty due to the $m$=1 DCP in the $X$ direction is estimated to be approximately $0.5\times10^{-16}$. The measured tilt sensitivity at the optimum power ($\pi/2$) in the $Y$ direction is $0.78\times10^{-15}$ mrad$^{-1}$. With the microwave power balanced within 20% and the optimal tilt angle with uncertainty of 0.44 mrad, the fractional frequency uncertainty due to the $m$=1 DCP in the $Y$ direction is estimated to be approximately $0.7\times10^{-16}$.

In summary, the total frequency correction for the DCP shift is 0, with the total uncertainty evaluated to be $0.87\times10^{-16}$. This uncertainty is derived from the square root of the quadratic sum of the uncertainties for the $m=0$ and $m=1$ DCP shifts in both $X$ and $Y$ directions.

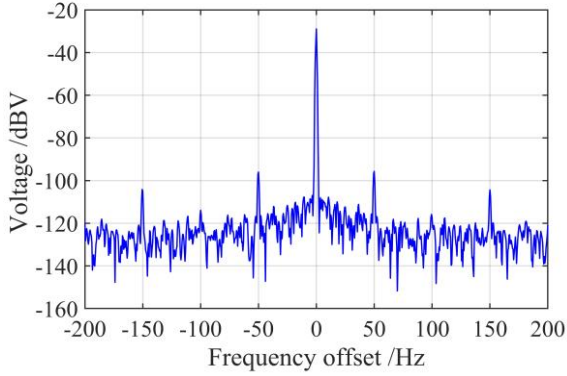

**Figure 12.** RF spectrum of the interrogation microwave, showing spurs at offsets of ±50 Hz and ±150 Hz from the clock frequency, measured to be less than 60 dB relative to the carrier power.

*C. Microwave spectrum impurity.* Spectral impurities in the Ramsey interrogation microwaves can introduce systematic frequency shifts [39]. These impurities may originate from the local oscillator or any active components in the microwave synthesizer, with the 50 Hz line frequency and its harmonics being the most significant noise sidebands. These sidebands lead to a frequency shift that depends on the sideband offset from the carrier, the sideband power, and their asymmetry [39, 40].

The spectrum of the interrogation microwave was obtained by mixing it with another microwave signal detuned 50 kHz away from the clock frequency, and shown in figure 12. Spurs at offsets of ±50 Hz and ±150 Hz from the clock frequency were measured to be less than 60 dB relative to the carrier power and were of equal amplitude to within 1 dB. Higher-order harmonics were found to be much smaller and negligible.

For a typical microwave interrogation duration $\tau_{in}$ of about 11 ms, a free evolution duration $T_R$ of about 515 ms, and a $\pi/2$ Ramsey pulse (setting $b\tau_{in}=\pi/2$ with an inaccuracy of approximately 10%) interrogation, the corresponding fractional frequency shift due to the microwave spectrum impurity is estimated to be $1.0\times10^{-17}$, according to the analysis in Reference [40].

*D. Overall microwave-power-related frequency shift.* After balancing both the phase and amplitude of the cavity's four feeds, the overall microwave-power-related shift was evaluated by operating the fountain clock NIM6 at the optimum power ($\pi/2$ pulse) and odd multiples of the optimum power ($3\pi/2$, $5\pi/2$, $7\pi/2$ and $9\pi/2$ pulses) alternately. The results are presented in figure 13. The relative frequency difference between the $\pi/2$ pulse and the $3\pi/2$ pulses was measured to be $(0.6\pm5.3)\times10^{-16}$ during two-day measurement. The results further support that the overall microwave-power-related frequency shift is small.

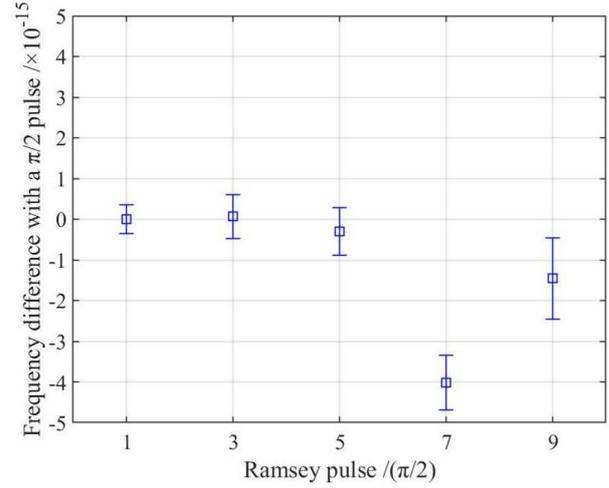

**Figure 13.** Fractional frequency shift of NIM6 as a function of Ramsey microwave amplitude. The relative frequency differences between the $\pi/2$ pulse and the $3\pi/2$, $5\pi/2$, $7\pi/2$, $9\pi/2$ pulses are measured to be $(0.6\pm5.3)\times10^{-16}$, $(-3.1\pm5.8)\times10^{-16}$, $(-4.0\pm0.7)\times10^{-15}$, $(-1.5\pm1.0)\times10^{-15}$, respectively.

*4.2.4 Blackbody radiation shift*

The blackbody radiation-induced frequency shift is caused by the electric field of ambient blackbody radiation. This shift can be calculated according to the equation [41]

$$\left(\frac{\Delta f}{f}\right)_{BBR} = \beta\left(\frac{T}{T_0}\right)^4\left[1+\varepsilon\left(\frac{T}{T_0}\right)^2\right] \qquad (9)$$

where $T_0=300$ K, $T$ is the effective temperature felt by the atoms during the Ramsey interrogation. The numerical values of the coefficients $\beta=-1.710(6)\times10^{-14}$ Hz and $\varepsilon=1.3(1)\times10^{-2}$ are taken from References [42] and [43], respectively. The Cs fountain NIM6 is located in a temperature-controlled laboratory where the ambient temperature varies by less than $\pm 0.2$ K from day to day. The flight tube is surrounded by an isothermal liner heat pipe, ensuring a more uniform and stable temperature along the atomic trajectory. Additionally, the flight tube is surrounded by a C-field solenoid coil mount and enclosed with three layers of magnetic shielding. These further reduce temperature variation in the flight tube zone compared to the ambient room temperature.

Two calibrated Pt thermometers were mounted at the top and bottom of the flight tube, with the recorded temperatures shown in figure 14. These thermometers have an uncertainty of less than 0.05 K and exhibit a very slow drift, which is verified using two additional Pt thermometers mounted outside the shield, at the top and bottom. The temperature difference between the top and bottom of the tube was measured to be less than 0.05 K, and the difference between the temperatures of the outer wall of the tube and that around the atom was less than 0.1 K. Taking an average temperature of the fight tube to be 296.78 K with an uncertainty of 0.2 K, the relative BBR shift for the Cs fountain NIM6 is calculated to be $-165.9 \times 10^{-16}$ with an uncertainty of $0.5 \times 10^{-16}$.

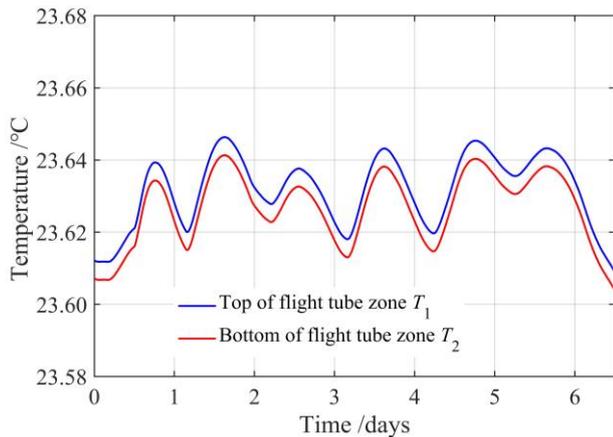

**Figure 14.** Temperature variations in the free drift zone of the fountain clock during the operating cycle, as measured by 2 Pt thermometers mounted at the top and bottom of the flight tube.

*4.2.5 Gravitational redshift*

The gravitational red shift is given by [44]

$$\left( \frac{\Delta f}{f} \right)_{\text{G}} = \frac{g}{c^2} \cdot H \approx 1.09 \times 10^{-16} \cdot H \qquad (10)$$

where $g$ is the local gravitational acceleration, $c$ is the speed of light, and $H$ is the time-averaged height of the atoms above the sea level during their Ramsey interrogation. In 2023, the NIM6 was relocated to a new building on the NIM Changping campus. Its time-averaged height $H$ has been precisely determined to be $(78.9 \pm 0.2)$ m, employing a method to that outlined in [45]. This measurement was conducted by the dimensional metrology laboratory of NIM, with the results traceable to a leveling benchmark situated within our campus. This benchmark, in turn, is referenced to the China 1985 national height datum, ensuring accuracy and reliability. The overall uncertainty in the orthometric height calculation encompasses both the leveling error and the potential vertical datum offset between the regional and global height datums, as discussed in [46]. Based on this measurement, the relative frequency shift due to the gravitational potential in NIM6 is calculated to be $86.0 \times 10^{-16}$ with an uncertainty of $0.2 \times 10^{-16}$.

*4.2.6 Light shift*

Any leaking near-resonant laser light interacting with atoms during the Ramsey interrogation time can induce a shift in the clock transitions due to the AC Starks effect [43]. This light shift depends on the light intensity and the detuning from atomic transitions. It is critical to block all the resonant light during the Ramsey interrogation to minimize this shift. Several mechanisms have been implemented in the fountain clock NIM6. First, the RF powers supplied to the AOMs (which control the cooling, detection and re-pumping beams) are switched off during the whole Ramsey interrogation duration. Second, mechanical shutters are used to further block the laser beams entering the optical fibers that deliver light to the fountain, as shown in figure 2. Third, the detection beam is detuned by -90 MHz to move it well away from the $F$=4 to $F'$=5 transition.

To evaluate the magnitude of the residual light shift, the NIM6 was operated in two alternating modes as described below. Mode 1 is a routine operation mode that all the above-mentioned mechanisms are used. In mode 2, the mechanical shutters used for detection and re-pumping beams remain open all the time, while all other conditions remain the same. The fractional frequency difference between the two modes was measured to be $(1.0 \pm 0.5) \times 10^{-15}$ over two-day measurement. Considering the mechanical shutters reduce the light intensity by at least three orders of magnitude [12], the uncertainty of the light shift in regular operation is estimated to be less than $1.0 \times 10^{-18}$. Therefore, no light shift correction is applied to the fountain frequency.

*4.2.7 Majorana transitions*

Majorana transitions occur when the magnetic field crosses zero or changes too rapidly along the atomic path after state-selection [47, 48]. Similar to the NIM5, Majorana transitions in the Cs fountain NIM6 are minimized to a negligible level by carefully shaping the magnetic field along the entire atomic trajectory, particularly near the endcaps of the magnetic shields. The adopted methods include that, the entire vacuum system of the NIM6 is wrapped with a layer of soft iron magnetic shield. Furthermore, an auxiliary coil, strategically positioned above the detection chamber (adjacent to the bottom endcap of the magnetic shields), instantaneously establishes a B field along the Z-axis after OM launching. This approach compensates for the field gradient induced by the 3-layers of μ-metal, ensuring a more uniform B field along the atom's trajectory and eliminating any potential zero-field crossings. Moreover, all magnetic fields generated by the compensation coils (in the Z-direction), C-field coils, and supplementary coils are aligned with the Earth's B field in the Z-direction and optimized to maximize transition probability. This was verified through the measurement of atom populations in the $|F$=3$\rangle$ states after state selection, as depicted in figure 15. After state selection, less than 1.4% of atoms remained in the $|F$=3, $m_F \neq 0\rangle$ states. The corresponding relative frequency shift due to Majorana transition is less than $1.0 \times 10^{-17}$, following the analysis of Reference [49].

*4.2.8 Rabi and Ramsey pulling*

The presence of other hyperfine transitions close to the clock transition can perturb the measured clock frequency due to Rabi and Ramsey pulling [50, 51]. Rabi-pulling mainly occurs when the microwave field drives neighboring $\Delta m_F=0$ transitions between magnetically sensitive states, with the dominant contribution coming from transitions between $|F=3, m_F=\pm1\rangle$ and $|F=4, m_F=\pm1\rangle$ states. The difference in the transition strengths between these states creates an asymmetry in the line shape of the $m_F=0$ clock transition, leading to a frequency shift. Whereas Ramsey pulling happens when the microwave field drives the $\Delta m_F=\pm1$ transitions, specifically $|F=3, m_F=0\rangle\leftrightarrow|F=4, m_F=\pm1\rangle$ and $|F=4, m_F=0\rangle\leftrightarrow|F=3, m_F=\pm1\rangle$. The asymmetry in the transition strengths induces an asymmetry in the Ramsey fringes, which also results in a frequency shift. Both Rabi and Ramsey pulling effects are therefore caused by asymmetries in the populations of the $m_F=\pm1$ states, leading to shifts in the clock frequency.

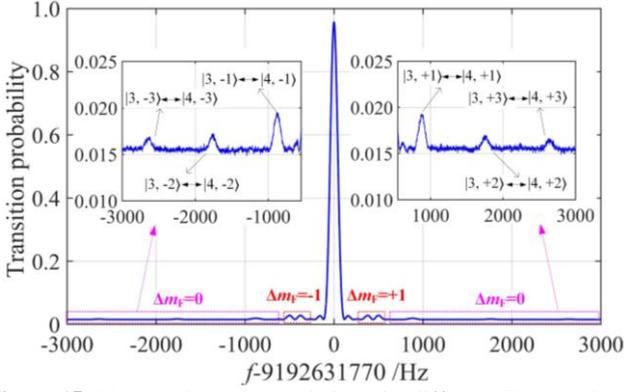

**Figure 15.** Measured atom populations in different Zeeman levels with Rabi signals. The frequency scanning step is 2.0 Hz, and the microwave power is set to $b\tau_{in}=\pi/2$. The $|F=3, m_F=\pm1\rangle\leftrightarrow|F=4, m_F=\pm1\rangle$ transitions are located at an offset frequency of $\pm877.4$ Hz.

In the fountain clock NIM6, as shown in figure 15, the measured populations of atoms in the $|3, m_F=\pm1\rangle$ states are about 0.8% of those in the $|3, 0\rangle$ state. The population imbalance $(p_{1,1}-p_{-1,1})/(p_{1,1}+p_{-1,1})$ between the $|3, m_F=\pm1\rangle$ states is determined to be about 2%, where $p_{1,1}$ and $p_{-1,1}$ represent the populations of the $|3, m_F=+1\rangle$ and $|3, m_F=-1\rangle$ states, respectively. Additionally, the asymmetry between the $|3, m_F=0\rangle\leftrightarrow|4, m_F=+1\rangle$ and $|3, m_F=0\rangle\leftrightarrow|4, m_F=-1\rangle$ transition probabilities is calculated to be $|(p_{0,+1}-p_{0,-1})/(p_{0,+1}+p_{0,-1})|=0.7\%$, where $p_{0,+1}$ and $p_{0,-1}$ are the probabilities for these respective transitions. Given the typical detuning of 877.4 Hz for the $\Delta m_F=0$ transitions $|3, \pm1\rangle\leftrightarrow|4, \pm1\rangle$ and 438.7 Hz for the $\Delta m_F=\pm1$ transitions $|3, 0\rangle\leftrightarrow|4, \pm1\rangle$ from the clock transition frequency, with a Rabi transition duration $\tau_{in}$ of about 11 ms and a free evolution duration $T_R$ of about 515 ms, the corresponding Rabi and Ramsey pulling frequency shifts are evaluated to be less than $1.0\times10^{-17}$ and $1.0\times10^{-18}$, respectively, using the theory outlined in [52] and [50]. These shifts are not corrected and are treated as uncertainty due to Rabi and Ramsey pulling.

### 4.2.9 Cavity pulling

The cavity pulling effect arises from the mistuning between the resonant frequency of the Ramsey cavity and the atomic transition frequency, as well as the interaction of atomic coherence with its own radiated field [53, 54]. This effect contributes to a frequency shift in two main ways. The first one (the first-order cavity pulling effect) is due to variations in the magnetic field phase inside the cavity, caused by its interaction with the magnetic moment of the atoms. The resulting frequency shift is proportional to the atom numbers within the cavity. In the Cs fountain NIM6, the first-order cavity pulling effect is typically small under normal operation conditions, due to the low atom number used. Additionally, this shift is continuously corrected along with the collisional shift by extrapolating to zero atomic density. The other one (the second-order cavity pulling effect) is related to the frequency dependence on the amplitude of the Ramsey cavity microwave field, and it is expressed by the following equation [53, 54]

$$\left(\frac{\Delta f}{f_0}\right)_{\text{cavity-pulling}} = \frac{\delta f_c}{f_0}\frac{8}{\pi^2}\frac{Q_c^2}{Q_{at}^2}\,b\tau_{in}\,\cot(b\tau_{in}) \quad (11)$$

where $\delta f_c=f_c-f_0$ is the cavity resonant frequency $f_c$ detuned from the atomic transition frequency $f_0$, $Q_c$ is the quality factor of the cavity. $Q_{at}=2f_0T_R$ is the quality factor of the atomic transition, $T_R$ is the effective Ramsey interrogation time. $b$ is the equivalent microwave magnetic field amplitude inside Ramsey cavity, and $\tau_{in}$ is the Rabi duration as atoms pass through the cavity.

During fountain clock operation, ambient temperature changes can shift the resonant frequency of the Ramsey cavity from its original design, leading to a second-order cavity pulling frequency shift [54]. For the Cs fountain clock NIM6, the Ramsey cavity $Q$-factor is measured to be about 12000, with a detuning $\delta f_c$ of about 10 kHz at lab temperature. The thermal coefficient of the cavity's resonant frequency is about -150 kHz/K, meaning that a temperature variation of 0.4 K ($\pm0.2$ K fluctuation in our lab) results in a 60 kHz shift in the cavity resonant frequency. During operation of the NIM6 with a $\pi/2$ Ramsey pulse (with a 10% inaccuracy in setting $b\tau_{in}=\pi/2$), the temperature fluctuation-induced second-order cavity pulling frequency shift is evaluated to be $2.2\times10^{-18}$. There, this frequency shift is sufficiently small that no correction is necessary.

### 4.2.10 Collisions with background gas

The background gas collisional shift is primarily caused by collisions between cold clock atoms with $H_2$ molecules, as well as room temperature background Cs atoms [55, 56]. In the Cs fountain NIM6, atoms are captured in the lower 3D-MOT chamber and launched to the upper OM chamber at a $10°$ angle (as shown in figure 1). The connections between the two chambers and between the OM chamber and the detection chamber are equipped with small tubes designed for differential pumping. Several graphite rings are placed inside the tube to getter ascending background Cs atoms. As a result, the Cs vapour level is significantly lower than that in the fountain NIM5 [12]. The primary source of background gas frequency shift in NIM6 is due to collisions with $H_2$ molecules. From the reading of the top ion pump connected to the flight tube, a background gas pressure less than $2.0 \times 10^{-8}$ Pa is observed. According to the pressure shift coefficients reported in the literature [57], the relative frequency shift due to background gas collisions is estimated to be less than $1.0 \times 10^{-17}$.

**Table 1.** Accuracy budget of the NIM6, listing physical effects, frequency corrections, and type-B uncertainty in a unit of $10^{-16}$.

| Physical effect | Bias /$10^{-16}$ | Uncertainty /$10^{-16}$ |
|---|---|---|
| Second-order Zeeman | 728.8 | 0.7 |
| Cold collisions | -22.0[a] | 1.7 |
| Microwave interferometric switch | 0.0 | 1.0 |
| Microwave leakage | 0.0 | 0.1 |
| DCP | 0.0 | 0.87 |
| Microwave spectral impurities | 0.0 | 0.1 |
| Blackbody radiation | -165.9 | 0.5 |
| Gravitational red shift | 86.0 | 0.2 |
| Light shift | 0.0 | 0.01 |
| Majorana transition | 0.0 | 0.1 |
| Rabi and Ramsey pulling | 0.0 | 0.1 |
| Cavity pulling | 0.0 | 0.02 |
| Collision with background gases | 0.0 | 0.1 |
| Total | 626.9[a] | 2.3 |

[a]The collisional shift is calculated at low density.

*4.2.11 Total accuracy budget of NIM6*

In summary, several other effects, such as the microwave lensing effect, relativistic Doppler effect, and others, may also contribute to shifts in the measured clock frequency. However, these effects are all well below $10^{-17}$ and are therefore negligible. Table 1 provides a summary of the biases and uncertainty contributions in the NIM6. The total type-B uncertainty is calculated to be $2.3 \times 10^{-16}$, with the largest contributions arising from cold collisions, the microwave interferometric switch, and the DCP effect.

## 5. Frequency comparison with UTC and PFSs

Along with its uncertainty evaluation, several frequency comparisons between the Cs fountain NIM6 and both TAI and PFSs ensemble have been conducted since June 2024. These comparisons were performed using the standard Time and Frequency Two-Way Satellite Transfer (TFTWST) method, where a local H-maser (BIPM code 1404821 (H21)), which contributes to UTC, was used as the frequency reference for the NIM6. Each measurement period lasts between 25 to 30 days. The averaged fractional frequency difference between the NIM6 and UTC is derived from the

**Table 2.** Comparisons of uncertainty budget for the NIM6. The $u_A$ is $2.3 \times 10^{-16}$ within the measurement time of 30 days (MJD 60519-60549).

| Effects | Relative frequency uncertainty |
|---|---|
| Stability $u_A$ | $2.3 \times 10^{-16}$ |
| Systematic $u_B$ | $2.3 \times 10^{-16}$ |
| Link to H-master $u_{\text{link/lab}}$ | $< 1.0 \times 10^{-16}$ |
| Link to TAI $u_{\text{link/TAI}}$ | $2.1 \times 10^{-16}$ |
| Final uncertainty into TAI | $4.0 \times 10^{-16}$ |

**Table 3.** Comparisons between the NIM6 and UTC(NIM), UTC and PFSs. The values are expressed in fractional units of $10^{-16}$.

| Measurement period/ MJD | $d$ (NIM6-UTC(NIM)) | $d$ (NIM6-UTC) | $d$ (NIM6-PFSs) | Uptime ratio /% |
|---|---|---|---|---|
| 60464-60489 | 2.3 | 2.8 | 3.1 | 99.4 |
| 60494-60519 | 1.8 | -3.2 | -3.4 | 98.7 |
| 60519-60549 | -4.8 | -3.2 | -3.8 | 99.1 |

data provided in the monthly Circular T reports published by the BIPM, using the following formula

$$y_M(\text{NIM6} - \text{TAI}) = y_M(\text{NIM6} - \text{H21}) +$$
$$y_M(\text{H21} - \text{UTC(NIM)}) - \quad (12)$$
$$y_M(\text{UTC} - \text{UTC(NIM)})$$

The combined uncertainty of the overall effects, including the links used for the comparison, is detailed in table 2. The total comparison uncertainty is calculated as the square root of the sum of the squares of the stability uncertainty $u_A$ and the systematic uncertainty $u_B$ of the NIM6 ($y_M$(NIM6-H21)), the link uncertainty $u_{\text{link/lab}}$ between the NIM6 and H21, and the frequency transfer uncertainty $u_{\text{link/TAI}}$ from H21 to TAI through the remote frequency comparison link. The link uncertainty $u_{\text{link/lab}}$ comprises two terms. The first term accounts for phase fluctuations caused by the cables connecting H21 to NIM6, which is estimated to be less than $1.0 \times 10^{-16}$. The second term arises from the dead time between the NIM6 measurements. Given that the dead time for all measurement remained below 2%, the corresponding uncertainty is considered negligible. The link uncertainty $u_{\text{link/TAI}}$ from H21 to TAI is derived from the time uncertainty between the UTC and UTC(NIM), as reported in the monthly Circular T reports.

Figure 16 shows the daily frequency data relative to two successive measurements taken between MJD 60464 to 60548. During this period, the NIM6 was alternately operated at high and low densities, with the frequency extrapolated to zero density. Between MJD 60464 to 60548, the fountain NIM6 was compared with UTC(NIM), UTC and the Cs fountain PFSs reported to the BIPM. The results of these comparisons, published in Circular T and summarized in table 3, support independently evaluated frequency shifts of NIM6 and the associated uncertainties.

## 6. Conclusions

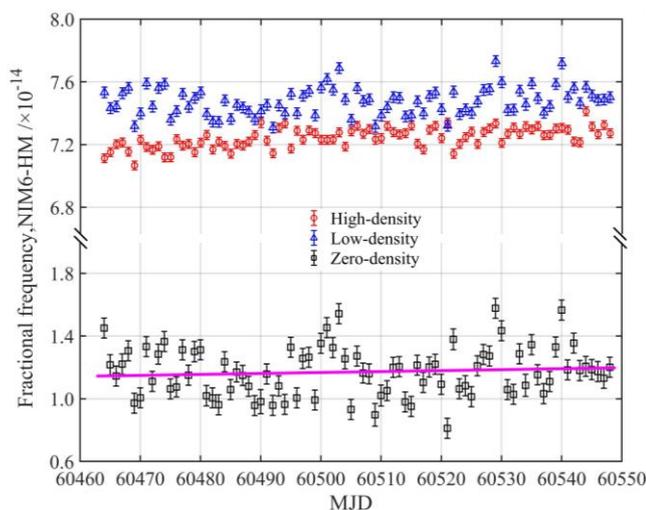

**Figure 16.** Frequency measurement of the fountain NIM6 versus H21. The slope of the fitted pink curve is $1.6 \times 10^{-17}$.

We present the first comprehensive uncertainty evaluation of the Cs fountain NIM6, a third-generation fountain primary frequency standard developed at the NIM in China. The clock demonstrates a short-term fractional frequency instability of $1.0 \times 10^{-13} \tau^{-1/2}$ when operated at high density, the short-term fractional frequency stability of averaging time. The overall type-B uncertainty is $2.3 \times 10^{-16}$. Comparisons with UTC(NIM), UTC and PFSs show an agreement with the evaluated frequency shifts and associated uncertainties. The clock NIM6 will function as a second-generation primary frequency standard for China and play a role in steering International Atomic Time (TAI).

### Acknowledgments

This work is supported by the National Natural Science Foundation of China Youth Science Foundation Project under Grant 12303074, the Key R&D Projects of the Ministry of Science and Technology under Grant 2021YFF0603800, the Fundamental Research Funds of NIM under Grant AKYJJ2402, and the Fundamental Research Funds for the Key Areas of National Institute of Metrology of China under Grant AKYZD2201-2.

The authors extend their special gratitude to Tiachu Li (1945-2022) for his significant and invaluable contributions to various stages of the development of the Cs fountain NIM6.

**Disclosures**. The authors declare no conflicts of interest.